\documentclass[twocolumn]{aastex63}
\usepackage{times}
\usepackage{amsmath}
\usepackage{textcomp}
\usepackage{amssymb}
\usepackage{natbib}
\usepackage{float}
\usepackage{color}
\usepackage{graphicx}
\newcommand{\Msun}{\ensuremath{M_{\odot}}}

\newcommand{\ergflux}{\mbox{${\rm \, erg \,\, cm^{-2} \, s^{-1}}$}}
\newcommand{\gm}{$\gamma$}

\received{\today}
\revised{\today}
\accepted{\today}
\submitjournal{ApJ}

\shorttitle{Gamma-ray Emitting Head-Tail Radio Galaxy ESO 137$-$G007}
\shortauthors{Paliya et al.}

\begin{document}
\title{A Gamma-ray Emitting Head-Tail Radio Galaxy Swimming Through the Hot Medium of the Merging Cluster Abell~3627}

\correspondingauthor{Vaidehi S. Paliya}
\email{vaidehi.s.paliya@gmail.com}

\author[0000-0001-7774-5308]{Vaidehi S. Paliya}
\affiliation{Inter-University Centre for Astronomy and Astrophysics (IUCAA), SPPU Campus, Pune 411007, India}

\author[0000-0002-8434-5692]{M. B{\"o}ttcher}
\affiliation{Centre for Space Research, North-West University, Potchefstroom, 2531, South Africa}

\author[0000-0002-5854-7426]{Swayamtrupta Panda}
\altaffiliation{Gemini Science Fellow and Visiting astronomer, Cerro Tololo Inter-American Observatory at NSF’s NOIRLab, which is managed by the Association of Universities for Research in Astronomy (AURA) under a cooperative agreement with the National Science Foundation.}
\affiliation{International Gemini Observatory/NSF NOIRLab, Casilla 603, La Serena, Chile}

\author[0000-0003-0841-7823]{Kiran Wani}
\affiliation{Indian Institute of Astrophysics, Block II, Koramangala, Bengaluru, Karnataka 560034, India}

\author[0000-0002-4464-8023]{D. J. Saikia}
\affiliation{Fakultat f\"ur Physik, Universit\"at Bielefeld, Postfach 100131, D-33501 Bielefeld, Germany}
\affiliation{Assam Don Bosco University, Guwahati 781017, Assam, India}

\author[0000-0002-9699-6257]{Mahadev Pandge}
\altaffiliation{DST-INSPIRE Faculty}
\affiliation{Dayanand Science College, Barshi Road, Latur 413512, Maharashtra, India}

\author[0000-0002-4998-1861]{C. S. Stalin}
\affiliation{Indian Institute of Astrophysics, Bengaluru, 560034, India}

\begin{abstract}
Head-tail radio sources are jetted active galactic nuclei (AGN) moving through the dense intracluster medium of galaxy clusters. They are extremely rare in the \gm-ray sky, likely because they are usually observed at large viewing angles. Therefore, every new \gm-ray detection of head-tail radio galaxies allows us to probe the origin of the high-energy emission in this unique class of AGN. Here we report, for the first time, the association of ESO~0137$-$G007 ($z=0.016$), a head-tail radio galaxy located at the outskirts of the merging cluster Abell 3627, with the \gm-ray source FL16Y~J1615.5$-$6034. The motion of the galaxy through the dense cluster environment and subsequent ram pressure and/or intracluster medium turbulence have led to the formation of a $>$500 kpc-long jet exhibiting a twisted, wiggly radio morphology. Its optical spectrum, obtained with the 4.1~m SOAR telescope, is devoid of emission lines, indicating that the underlying accretion activity is radiatively inefficient. From the measured line-of-sight stellar velocity dispersion, we estimate the mass of the central supermassive black hole to be $(3.36\pm1.32)\times10^9$ \Msun. We briefly discuss several radiative models to explain the observed broadband spectral energy distribution of ESO~0137$-$G007. We conclude that deeper multi-wavelength observations of this unique \gm-ray emitting head-tail radio galaxy will enable us to better understand the interaction of the jet with the hot cluster environment, thus setting the stage for the research of this class of AGN with the upcoming Square Kilometer Array.

\end{abstract}


\section{Introduction}
High-energy observations of astrophysical sources reveal some of the most intriguing sites where matter and radiation interact. Given the large-scale environments of galaxy clusters, such observations provide tantalizing clues about the dynamics of the cluster gas and its interactions with member galaxies. Some of them exhibit spectacular radio jets which strongly affect the ambient environment \citep[cf.][]{2020MNRAS.496.2613B,2023ApJ...948..101S}. The radio emitting plasma is often shaped by the motion of the host galaxy and the ram pressure and/or the large-scale turbulence of the intracluster medium (ICM), leading to the formation of so-called head-tail radio sources \citep[e.g.,][]{1968MNRAS.138....1R,1979ApJ...234..818J,2017AJ....154..169S,2020MNRAS.493.3811S}. A small number of head-tail radio galaxies, e.g., IC~310 ($z=0.019$), have also been detected at MeV$-$GeV energies, indicating their jets to be able to accelerate particles to very high energies \citep[][]{2017A&A...603A..25A,2024ApJ...976..120P,2025ApJ...989...36P}. Since these objects are usually observed at large viewing angles, their rare \gm-ray detection makes them a primary target to study the high-energy radiative processes from a different vantage point than the more common beamed jetted active galactic nuclei (AGN) or blazars \citep[see, e.g.,][]{2022ApJ...931..138F,2024ApJ...965..163Y}. Recent numerical simulations, involving the evolution of cosmic-ray particle spectra with energy losses and stochastic turbulence acceleration in the jet-wind interaction, also indicate that they are extremely faint \gm-ray emitters \citep[][]{2023ApJ...951...76O}. However, with more than 17 years of almost uninterrupted monitoring of the \gm-ray sky by the Fermi Large Area Telescope (LAT), the potential to identify elusive \gm-ray-emitting head-tail radio galaxies remains high.

FL16Y~J1615.5$-$6034 is a \gm-ray emitting source with so far no known associated low-frequency counterpart \citep[][]{2022ApJS..260...53A,2026arXiv260222148B}. In this paper, we provide observational evidence for the association of FL16Y~J1615.5$-$6034 with the head-tail radio galaxy, ESO~0137$-$G007 ($z=0.016$) located at the outskirts of the merging cluster Abell 3627 \citep[][]{1989ApJS...70....1A}. ESO~0137$-$G007 exhibits an enigmatic $>$500 kpc long radio jet with wiggled and twisted morphology, and an arc-shaped radio filament beyond and mostly orthogonal to the collimated inner tail end as revealed by Australian Square Kilometer Array Pathfinder (ASKAP) data from the Evolutionary Map of the Universe \citep[EMU; Figure~\ref{fig:Fig1}; see][for details]{2024MNRAS.533..608K}. The source is traveling eastwards close to the plane of the sky, as revealed by the projected radio tail morphology and extent \citep{2024MNRAS.533..608K}. From X-ray imaging, it is evident that ESO~0137$-$G007 plunges through the dense ICM of Abell 3627, and the resulting ram pressure and shear forces sculpt its radio jets (Figure~\ref{fig:Fig1}, top left panel). Recently, \citet[][]{2026arXiv260312082G} reported the detection of a merger shock (Mach number $\sim$1.3) towards the tail-end of the helical jet. All in all, ESO~0137$-$G007 provides us a unique opportunity to explore the origin of the high-energy emission in what is likely to be a rare \gm-ray emitting head-tail galaxy moving in the dense and hot ICM of a merging cluster.

Throughout, we have adopted the flux density $F_{\nu}\propto\nu^{\alpha}$, where $\alpha$ is the radio spectral index. A flat cosmology with $H_0 = 70~{\rm km~s^{-1}~Mpc^{-1}}$ and $\Omega_{\rm M} = 0.3$ was used.

\section{Multiwavelength Association}\label{sec2}
\begin{figure*}
\hbox{
    \includegraphics[scale=0.26]{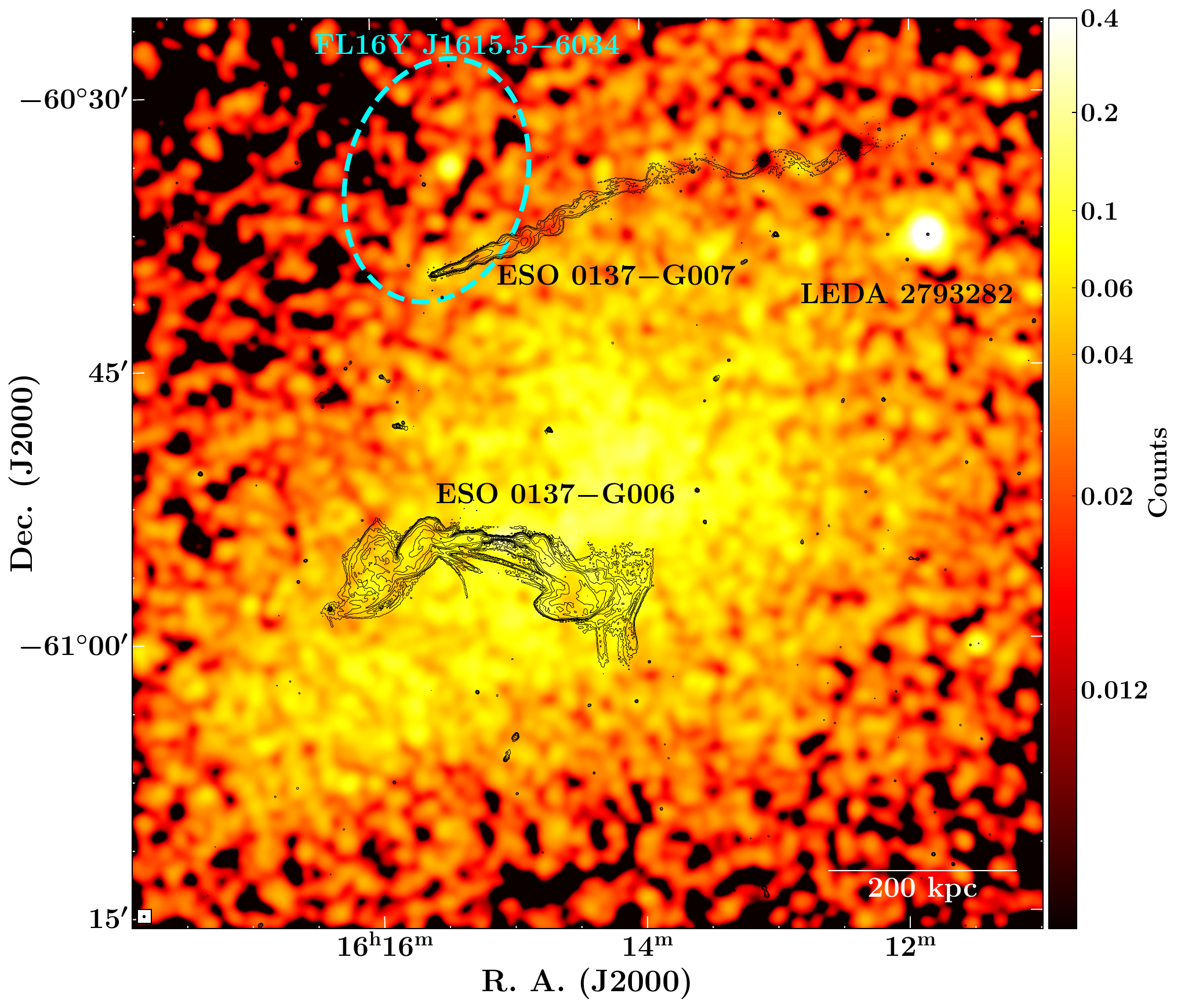}
   \includegraphics[scale=0.38]{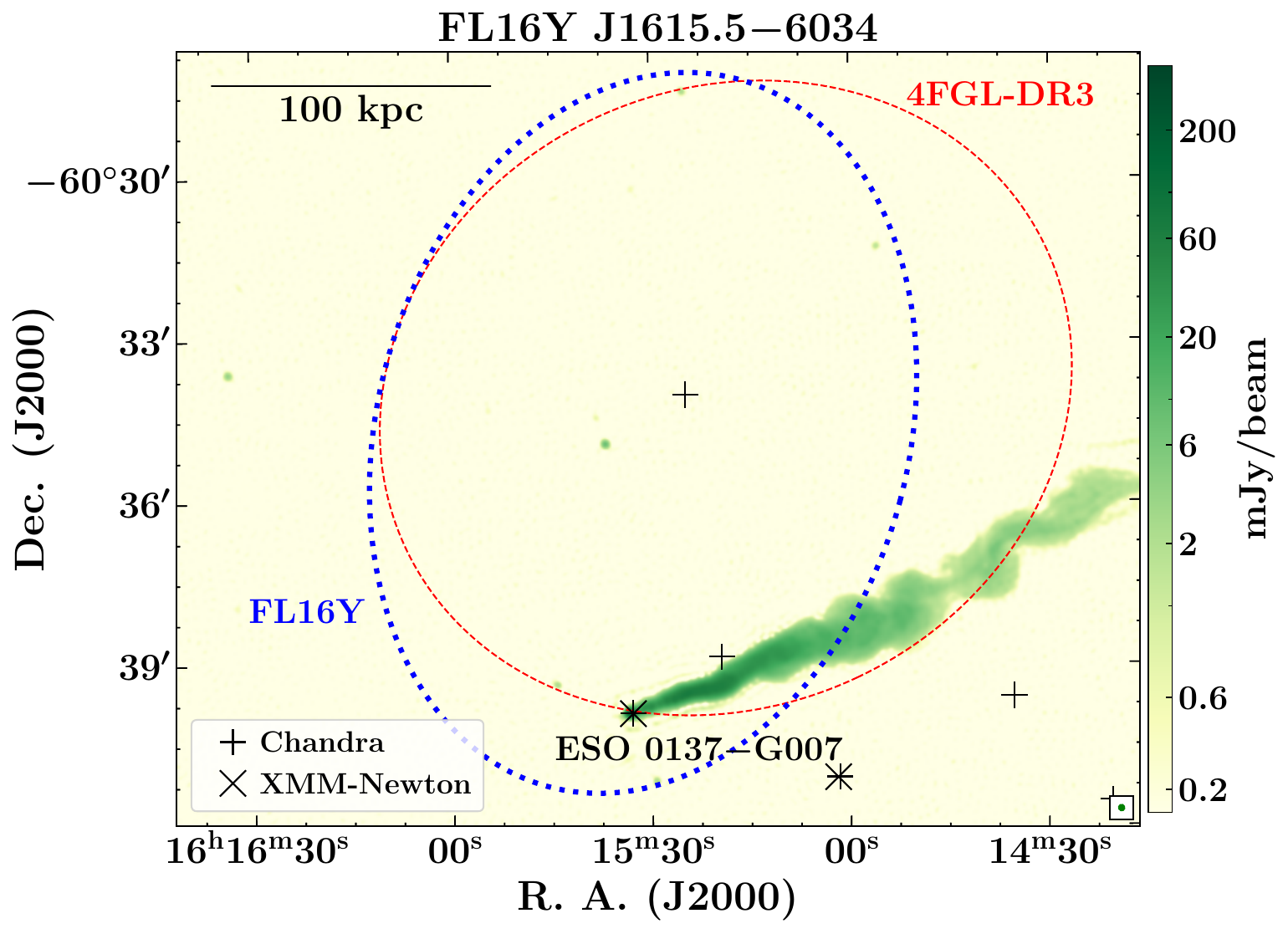}
    }
\hbox{\hspace{1.5cm}
   \includegraphics[scale=0.4]{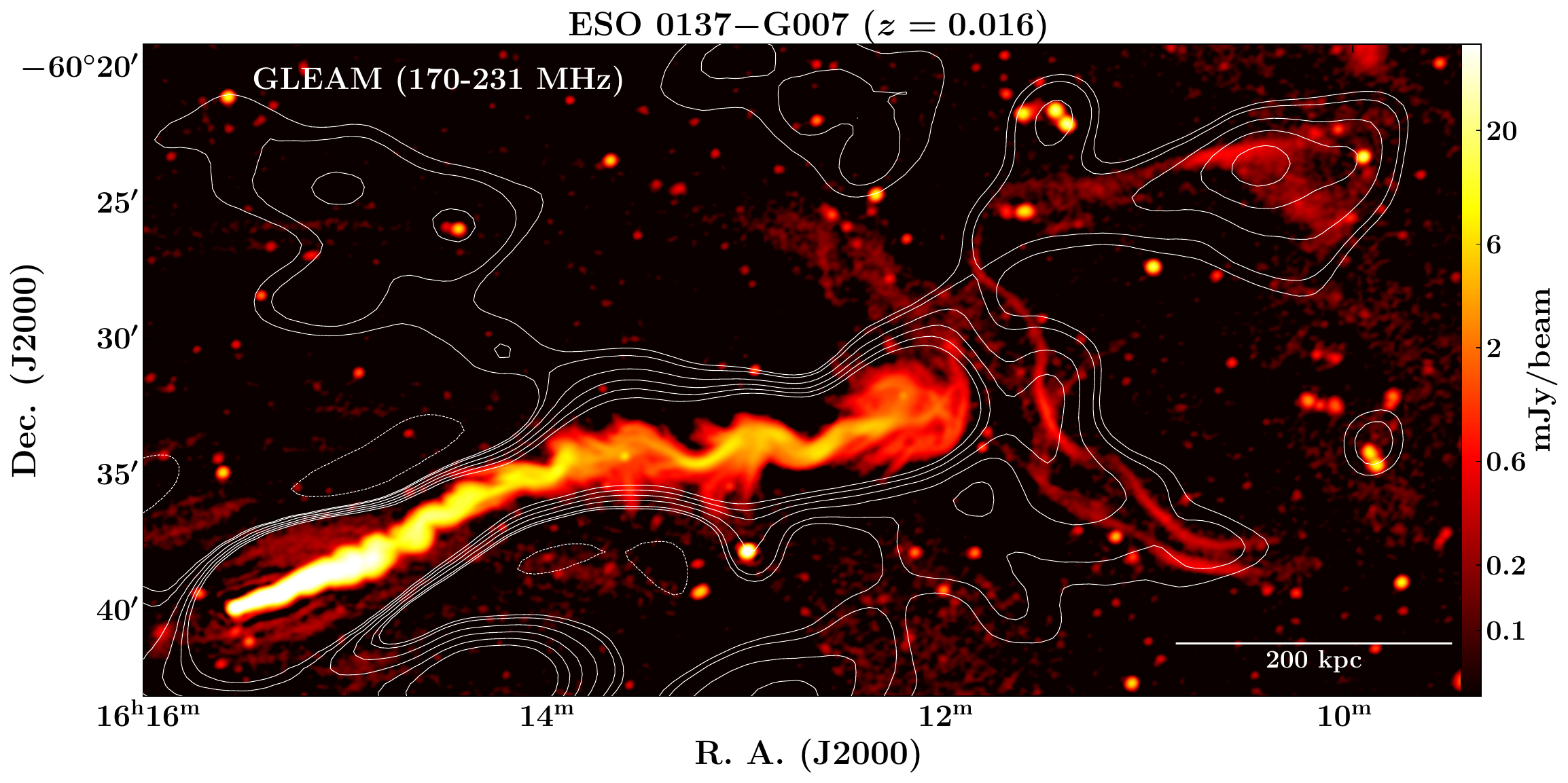}
}
\caption{Top Left: The 0.2$-$2.3 keV image of the merging cluster Abell 3627 taken with the eROSITA satellite \citep[][]{2024A&A...682A..34M}. The image is smoothed to highlight the faint diffuse features. Overplotted are the radio contours representing the ASKAP-EMU image. Top Right: False-color image of the ASKAP-EMU observation taken at 943 MHz. The plotted ellipses refer to the 95\% positional uncertainty in the optimized \gm-ray position as provided by the 4FGL-DR3 (red solid ellipse) and FL16Y (blue dotted ellipse) catalogs. The positions of the identified X-ray point sources are also highlighted, as labeled. Bottom: The 170$-$231~MHz-band contours made using the GaLactic and Extragalactic All-sky MWA (GLEAM) survey \citep{2015PASA...32...25W} superposed on the ASKAP-EMU image. GLEAM contours are plotted at 0.1, 0.2, 0.3, 0.4, 0.5, and 0.6 Jy beam$^{-1}$. The image is smoothed to highlight the faint features.}\label{fig:Fig1}
\end{figure*}

FL16Y~J1615.5$-$6034 was first reported as a \gm-ray source detected at 4.2$\sigma$ confidence level in the third data release of the fourth catalog of the Fermi-LAT detected \gm-ray sources \citep[4FGL-DR3, named as 4FGL~J1615.3$-$6034;][]{2022ApJS..260...53A}. In 4FGL-DR4 and the latest Fermi-LAT 16-year Source List (FL16Y), it was reported to be detected with 4.46$\sigma$ and 5.16$\sigma$ confidence levels, respectively \citep[][]{2023arXiv230712546B,2026arXiv260222148B}. The 0.1$-$100 GeV energy flux given in the 4FGL-DR3, 4FGL-DR4, and FL16Y catalogs are $(3.45\pm0.84)\times10^{-12}$ \ergflux, $(3.79\pm0.89)\times10^{-12}$ \ergflux, and $(3.42\pm0.77)\times10^{-12}$ \ergflux, respectively. The corresponding power-law photon index values are $2.63\pm0.15$, $2.67\pm0.14$, and $2.58\pm0.14$, respectively. These parameters indicate that FL16Y~J1615.5$-$6034 is a faint but persistent \gm-ray emitter with a soft spectrum. In the top right panel of Figure~\ref{fig:Fig1}, we show the radio image from ASKAP data (800$-$1088 MHz) centered at the \gm-ray source position and overplot the 95\% positional uncertainty ellipses from the 4FGL-DR3 and FL16Y catalogs. While the head-tail galaxy ESO 0137$-$G007 was barely at the border of the 4FGL-DR3 uncertainty ellipse, it lies well inside the uncertainty ellipse reported in the FL16Y catalog. The FL16Y catalog covered a larger data set (16 years) than 4FGL-DR3 (12 years), and thus has deeper sensitivity, suggesting that ESO~0137$-$G007 is potentially associated with FL16Y~J1615.5$-$6034. The \gm-ray spectral parameters FL16Y~J1615.5$-$6034 are consistent with those found for other \gm-ray detected misaligned jetted AGN \citep[][]{2024ApJ...976..120P}, thus providing further supportive evidence for the association.

We explored the possibility of a blazar or other AGN within the 95\% uncertainty region of FL16Y~J1615.5$-$6034 as the low-frequency counterpart of the \gm-ray object. We found 3 X-ray point sources considering the objects detected at $>$5$\sigma$ confidence level in the Chandra Source Catalog and XMM-Newton serendipitous source catalog \citep[][]{2020A&A...641A.136W,2024ApJS..274...22E}. The brightest of them (4XMM~J161532.9$-$603955, $F_{\rm 0.2-12~keV}=(2.50\pm0.47)\times10^{-13}$ \ergflux) is positionally consistent with ESO 0137$-$G007. The other two fainter X-ray sources, 2CXO~J161519.3$-$603852 ($F_{\rm 0.5-7~keV}=7.66^{+0.80}_{-0.79}\times10^{-14}$ \ergflux) and 2CXO J161524.6$-$603401 ($F_{\rm 0.5-7~keV}=3.17^{+0.52}_{-0.52}\times10^{-14}$ \ergflux), are likely to be low-mass stars \citep[][]{2024ApJ...971..180Y}. Indeed, neither of them is detected in the radio band as revealed by the ASKAP observations (Figure~\ref{fig:Fig1}, top right panel). A radio point source lies within the \gm-ray uncertainty ellipse at right ascension = 243$^{\circ}$.90316 and declination = $-$60$^{\circ}$.58241. However, it is considerably fainter ($F_{\rm 1.65~GHz}\approx11$ mJy) than ESO 0137$-$G007 (core $F_{\rm 1.65~GHz}\approx292$ mJy). It is also not detected by the Chandra/XMM-Newton satellites, thus unlikely to be the counterpart of FL16Y~J1615.5$-$6034. Therefore, ESO~0137$-$G007 is the most promising counterpart of the \gm-ray source.

We further quantified the association of ESO 0137$-$G007 with FL16Y~J1615.5$-$6034 by estimating the multiwavelength likelihood ratio ($LR$) and computing the Bayesian association probability based on the angular separation from the \gm-ray centroid and the surface density of comparable radio and X-ray sources \citep[e.g.,][]{1992MNRAS.259..413S,2018MNRAS.473.4937S}. We considered the eROSITA DR1 source catalog \citep[][]{2024A&A...682A..34M} and Rapid ASKAP Continuum Survey (RACS) source catalog \citep[][]{2024PASA...41....3D} to derive per-candidate $LR_k$ as follows:

\begin{equation}
    LR_k = \frac{f_k(\Delta x,\Delta y)}{\rho_s(k)},
    \label{eq1}
\end{equation}

where $f_k(\Delta x,\Delta y)$ is the spatial Gaussian probability density evaluated at the position of the $k^{th}$ counterpart within the Fermi-LAT error ellipse. The parameter $\rho_s(k)$ is the background number density defined as 
\begin{equation}
\rho_{\rm s}=\frac{n_{\rm s}}{A},~s\in\{\rm joint,~X-ray~only, Radio~only\}    
\end{equation}
for all radio and/or X-ray detected objects lying within the background annulus area $A$. For a given X-ray ($F_{\rm X}$) and radio ($F_{\rm R}$) flux thresholds, we have
\begin{equation}
    n_{\text{pairs}} = \#\{(i,j) : S_{X,i} \geq F_X, \, S_{R,j} \geq F_R, \, \theta_{ij} \leq r_{\text{match}}\}
\end{equation}
and
\begin{equation}
    n_{\text{X/R-only}} = {} \# \{i : S_{X/R,i} \ge F_{X/R}\} - n_{\text{pairs}}
\end{equation}
where $\theta_{ij}$ is the angular separation between the X-ray source $i$ and radio source $j$, and $r_{\rm match}$ is the cross-match radius chosen as 3 arcsec. The joint calculation used the \gm-ray source positional likelihood only once and considered the local density of RACS and eROSITA cross-matched sources satisfying both flux criteria. For ESO 0137$-$G007, we obtained $f(\Delta x,\Delta y)=1.28\times10^{-6}$ arcsec$^{-2}$. The background number density was computed within an annulus of inner and outer radii of 0$^{\circ}$.1 and 2$^{\circ}$ centered at the Fermi-LAT source. Taking the radio and X-ray brightness of ESO 0137$-$G007 as 292 mJy and $2.5\times10^{-13}$ \ergflux, respectively, we derived $\rho_{\rm s}=6.15\times10^{-9}$ arcsec$^{-2}$. The normalized separation of the \gm-ray source centroid with the RACS and eROSITA positions was found to be 1.91. Inserting these numbers in Equation~\ref{eq1}, we estimated the likelihood ratio to be $\mathcal{LR}_{\rm ESO}=208.7$. We also considered X-ray and radio sources, other than ESO 0137$-$G007, located within the 95\% Fermi-LAT uncertainty region. Since the detected X-ray sources have no radio counterparts and vice versa, we conservatively adopted 3$\sigma$ flux upper limits and derived the $LR$ values of 0.51 (5.5) for only X-ray (radio) detected objects.

Given $N$ mutually exclusive candidates $k=1,...,N$ competing for the same Fermi-LAT source, and assuming a single shared prior $Q$ that a true counterpart exists among them at all, the joint multi-candidate posterior reliability can be derived as follows \citep[][]{1992MNRAS.259..413S}:
\begin{equation}
    P_k = \frac{LR_k}{\sum_{j=1}^{N} LR_j + \frac{1 - Q}{Q}};~P_{\text{none}} = \frac{\frac{1 - Q}{Q}}{\sum_{j=1}^{N} LR_j + \frac{1 - Q}{Q}}.
\end{equation}
Here, $P_{\rm none}$ denotes the probability that none of the considered multi-wavelength sources are true counterpart of the \gm-ray emitter, thus setting $\sum_{k=1}^{N} P_k + P_{\text{none}} = 1
$. Assuming $Q=0.05$ \citep[cf.][]{2011ApJ...743..171A}, we estimated the association probability of ESO 0137$-$G007 being the low-frequency counterpart of FL16Y~J1615.5$-$6034 to be $89.3\%$, which is well-above of 80\% adopted in the Fermi-LAT catalogs \citep[][]{2011ApJ...743..171A}. For the sources detected only in the radio and X-ray bands, we estimated the association probabilities to be 2.35\% and 0.22\%, respectively. On the other hand, $P_{\rm none}$ was dervied to be 8.13\%. Therefore, comparing the association probabilities of all sources, we can conclude ESO 0137$-$G007 as the most plausible multiwavelength counterpart of FL16Y~J1615.5$-$6034.

\begin{figure*}
\hbox{
    \includegraphics[scale=0.5]{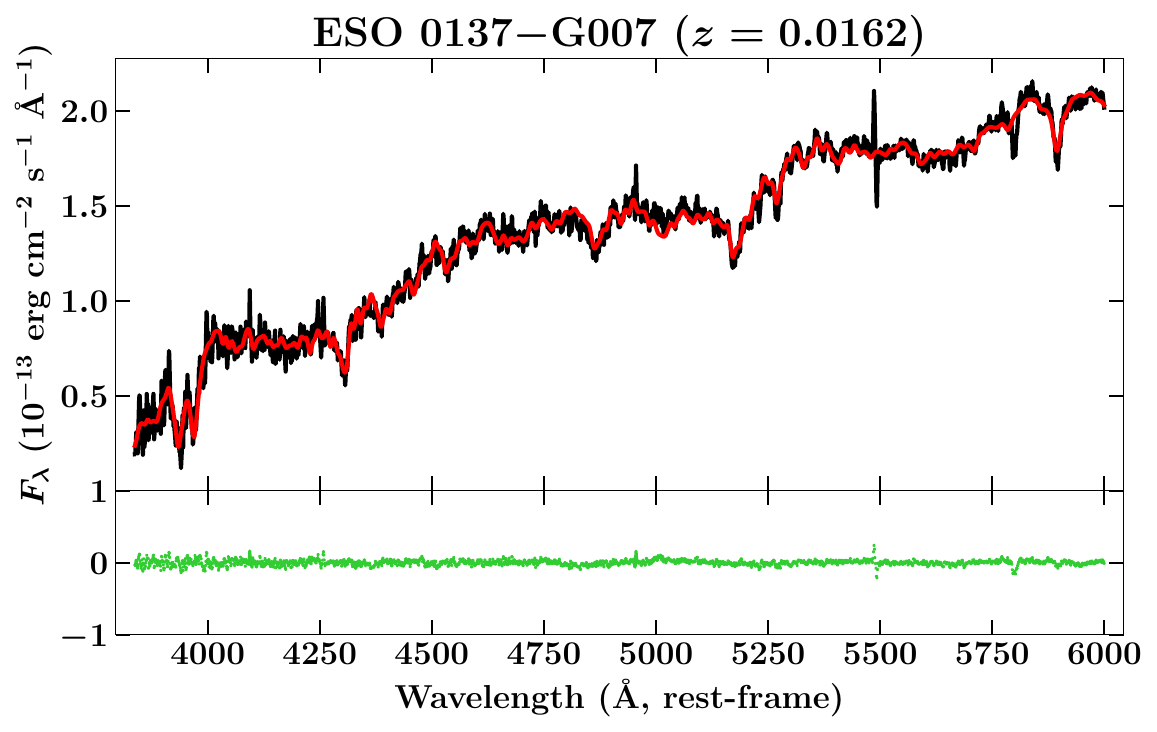}
   \includegraphics[scale=0.44]{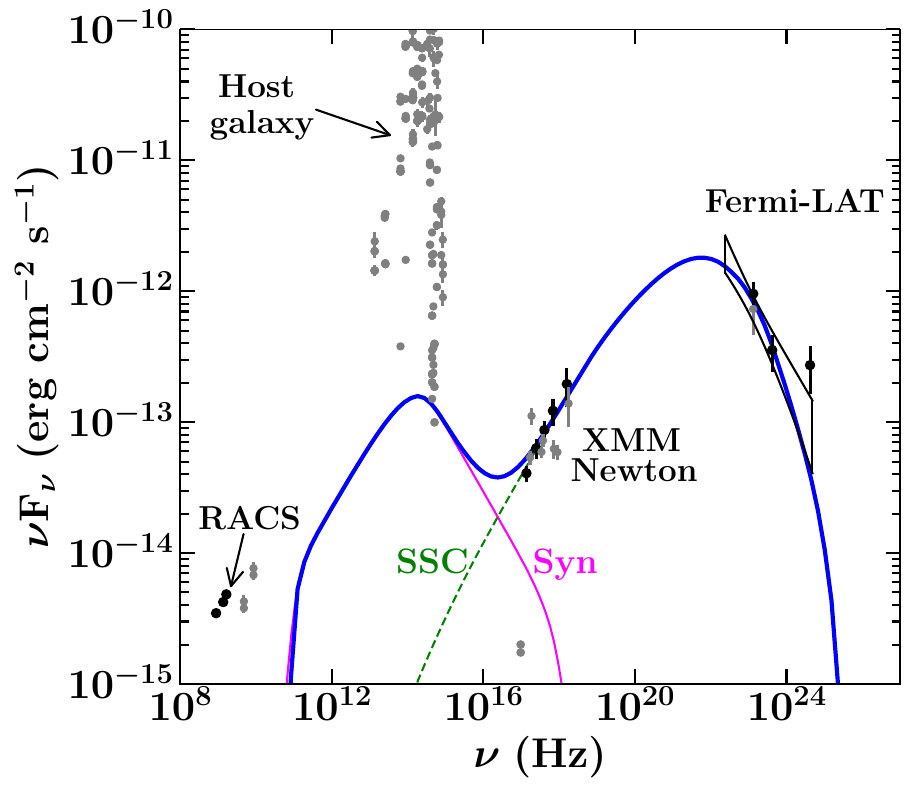}
    }
\caption{Left: Optical spectrum of ESO~0137$-$G007 taken with the SOAR telescope. The fitted stellar population synthesis model is shown with the red line. Right: The broadband SED of the head-tail radio galaxy reproduced with a stationary single-zone leptonic synchrotron-SSC radiative model. See the text for details.}\label{fig:Fig2}
\end{figure*}

\section{Central Engine and Ambient Environment}\label{sec3}
We procured a new optical spectrum of ESO~0137$-$G007 using the Goodman High-Transmission Spectrograph \citep{2004SPIE.5492..331C} mounted on the 4.1-meter Southern Astrophysical Research Telescope on 2025 May 10 (Program ID: SO2025A-020; co-PI: Panda). We used the M1 Red Camera (400 lines/mm, 0.45 arcseconds per pixel) and took 3 exposures of 600 seconds each, providing a resolving power of $R \sim 2000$ at 5000~\AA\ and covering a wavelength range of $3900$--$7000$~\AA. The detector was operated with a read noise of 3.45~$e^{-}$ and a gain of 1.54~$e^{-}$/ADU in $1\times1$ binning mode. Observations were conducted at an average airmass of 1.17 under stable atmospheric conditions.

The data reduction was performed following the standard procedures in Image Reduction and Analysis Facility \citep{1986SPIE..627..733T}. For the flux calibration, we observed the standard star EG~274. In the left panel of Figure~\ref{fig:Fig2}, we show the observed spectrum, which appears to be consistent with that typically observed from massive elliptical galaxies. No emission lines were detected, implying that the underlying accretion activity is radiatively inefficient. Since the host galaxy absorption features dominate the observed spectrum, we applied the penalized pixel fitting ({\tt pPXF}) software to determine the line-of-sight stellar velocity dispersion \citep[$\sigma^*$; see,][for details]{2004PASP..116..138C,2023MNRAS.526.3273C}. The fitted model is shown in the left panel of Figure~\ref{fig:Fig2}, and the $\sigma^*$ was estimated to be $429.4\pm14.8$ km~s$^{-1}$. Considering the commonly adopted $M-\sigma^*$ relation \citep[e.g.,][]{2009ApJ...698..198G}, we derived the mass of the central supermassive black hole as $(3.36\pm1.32)\times10^9$ \Msun.

ESO 0137$-$G007 lies $\sim$0.3 Mpc (projected) from the center of Abell~3627, a very massive ($M\sim10^{15}M_\odot$), X-ray–bright cluster at $z=0.0157$ \citep[][]{1996Natur.379..519K}. Being located close to the Galactic plane, Abell~3627 is highly obscured. Galaxy velocities in the Abell~3627 cluster have a dispersion of order 900 km s$^{-1}$, indicating it is dynamically evolved but hosts substructure.

\section{The Origin of the Broadband Radiation}
High-resolution radio observations of head-tail sources suggest that their inner jets are usually not affected by strong relativistic effects, although there are exceptions, as in the case of IC~310 which has also been detected in \gm-rays \citep[][and references therein]{2017A&A...608A..58T,2020MNRAS.499.5791G}. Important characteristics of relativistically beamed emission are (i) a prominent nuclear or core component with a flat radio spectrum, and (ii) significant flux density variability of this component. IC~310 has a flat radio spectrum with $\alpha_{365}^{4850}\sim -0.36$ and its radio emission is dominated by the radio core. From observations at 5~GHz with the Very Large Array, \citet[][]{2010MNRAS.404..180D} reported the peak brightness, which is from the radio core, to be 136 mJy beam$^{-1}$ and its total flux density to be 198 mJy, yielding a core dominance\footnote{The core dominance is defined as the ratio of the rest-frame core to extended flux densities at 3~GHz \citep[see. e.g.,][]{2024ApJ...976..120P}.} of $\sim$1.5. IC~310 has exhibited significant X-ray and \gm-ray flux variability and has also been found to be variable at radio frequencies \citep[][]{2014A&A...563A..91A,  2017A&A...603A..25A, 2020MNRAS.499.5791G}. On VLBI scales it has a core-jet structure similar to blazars \citep[cf.][]{2012A&A...538L...1K}. On the other hand, the radio spectrum near the core of ESO 0137$-$G007 is steep \citep[][]{2024MNRAS.533..608K}. We also calculated the core dominance using RACS-low data, and found it to be 0.05. This is also likely to be an upper limit as the steep spectrum of the core indicates inclusion of more extended emission in the core flux density. Such a low value of core dominance indicates weak beaming effects, unlike in the case of blazars \citep[][]{2001MNRAS.326.1455M,2015MNRAS.451.4193C,2024ApJ...976..120P}. To probe variability of the core in ESO 0137$-$G007 higher-resolution observations at radio frequencies are required to isolate the core, and higher signal-to-noise data at \gm-rays.

Moreover, ESO 0137$-$G007 does not lie in a region occupied by blazars in the Wide-field Infrared Survey Explorer color-color diagram \citep[also known as  WISE blazar strip;][]{2011ApJ...740L..48M}. Therefore, available observations do not support the possible beamed nature of the inner radio jet. High-resolution radio observations of this intriguing object will be needed to isolate the core and probe its small-scale structure.

\subsection{Broadband Modeling}
Given the orientation of the jet approximately perpendicular to our line of sight, radiation models not relying on relativistic Doppler boosting should be considered. The simplest case of such a model is a leptonic single-zone synchrotron-self-Compton model with a stationary emission region. As the infrared -- optical -- ultraviolet spectrum of the source is dominated by the host galaxy, the synchrotron peak is very poorly constrained. We scale our following estimates by a location at $\nu_{\rm sy} \equiv 10^{14} \, \nu_{\rm sy, 14}$~Hz and $\nu F_{\nu}^{\rm sy} \equiv 10^{-13} \, F_{\rm sy, -13}$~erg~cm$^{-2}$~s$^{-1}$. For an SSC peak at  $\nu_{\rm SSC} = 10^{21} \, \nu_{\rm SSC, 21}$~Hz, the peak of the radiating electron spectrum is required to be at $\gamma_{\rm e, pk} \sim 3 \times 10^3 \, (\nu_{\rm SSC, 21} / \nu_{\rm sy, 14})^{1/2}$. The peak synchrotron flux provides an estimate of the synchrotron photon energy density in the emission region of $u'_{\rm sy} \sim 4 \times 10^{-7} \, R_{\rm b, pc}^{-2}$~erg~cm$^{-3}$, where the  radius of the spherical emission region is parameterized as $R_b \equiv 1 \,  R_{\rm b, pc}$~pc. For an SSC peak flux of $\nu F_{\nu}^{\rm SSC} \sim 2 \times 10^{-12}$~erg~cm$^{-2}$~s$^{-1}$, the ratio of SSC to synchrotron peak fluxes constrains the magnetic field through the identity $u'_{\rm sy} / u'_{\rm SSC} = \nu F_{\nu}^{\rm sy}/ \nu F_{\nu}^{\rm SSC}$ to be $B \sim 7 \times 10^{-4} \, R_{\rm b, pc}^{-1}$~G. Finally, to produce a synchrotron peak at $\nu_{\rm sy} \sim 4 \times 10^6 \, B_G \, \gamma_{\rm e, pk}^2 \, {\rm Hz} \, \sim 3 \times 10^{10} \, (\nu_{\rm SSC, 21} / \nu_{\rm sy, 14}) \, R_{\rm b, pc}^{-1}$~Hz~$\sim 10^{14}$~Hz, we require $R_{\rm b, pc} \sim 3 \times 10^{-4}$, i.e., a very small emission region of radius $R_b \sim 10^{15}$~cm, which then implies a magnetic field strength of $B \sim 2$~G. 

In order to test the viability of a stationary SSC model with parmeters as estimated above, we employ the leptonic blazar emission model of \cite{Boettcher2013}, with bulk  Lorentz factor $\Gamma = 1.1$ and viewing angle $\theta_{\rm obs} = 80^{\circ}$, i.e., negligible Doppler boosting. We caution that the choice of a viewing angle of $80^{\circ}$ is a model assumption and not an observationally established value. Moreover, given the poor quality data, especially in the infrared--ultraviolet region, the model used here should be considered as an illustrative SSC realization rather than a statistically constrained fit. The code assumes continuous injection of a non-thermal, relativistic electron distribution, described by a power-law with low- and high-energy cut-offs at $\gamma_{\rm min/max}$, respectively. A tangled magnetic field is assumed, and for the SSC case, external target radiation fields for Compton scattering are neglected. The code evaluates an equilibrium  between the above-mentioned particle injection,  radiative  (synchrotron + SSC) cooling, and escape on a time scale $t_{\rm esc} = \eta_{\rm esc} \, R_b/c$. Specific parameters employed are listed in Table \ref{tab:SSCparameters}, and the resulting SSC model fit is shown in in the right panel of Figure~\ref{fig:Fig2}. 

\begin{table*}[htp]
\centering
\caption{Parameters for the stationary leptonic SSC fit to the SED shown in Figure~\ref{fig:Fig2}.
}
\label{tab:SSCparameters}

\begin{tabular}{lcc} 
 \hline
 Parameter & Symbol & Value \\
 \hline
 Minimum electron Lorentz factor  & $\gamma_{\rm e, min}$  & $4.7 \times 10^3$  \\
  Maximum electron Lorentz factor  & $\gamma_{\rm e, max}$  & $3.0 \times 10^5$  \\
  Electron spectral index  & $q_e$  & 3.0  \\
  Escape timescale parameter  & $\eta_{\rm esc}$  & 100  \\
  Magnetic field  & $B$  & 2~G \\
  Bulk Lorentz factor  & $\Gamma$  & 1.1  \\
  Viewing angle & $\theta_{\rm obs}$ & 80$^{\circ}$ \\
  Blob radius  & $R$  & $1.0 \times 10^{15}$~cm  \\
  \hline
         DERIVED QUANTITIES \\
  \hline
  Electron luminosity & $L_{\rm e}$ & $4.3  \times 10^{42}$~erg~s$^{-1}$  \\
  Magnetic field luminosity & $L_{\rm B}$ & $1.8 \times 10^{40}$~erg~s$^{-1}$  \\
  Ratio of magnetic field to electron energy densities & $u'_{\rm B}/u'_{\rm e}$ & $4.2 \times 10^{-3}$ \\
  \hline
\end{tabular}
\end{table*}

The parameters listed in Table \ref{tab:SSCparameters} appear reasonable and consistent with electron energized by shock acceleration in a small region in which the jet interacts with its external environment. 
The small emission-region size required suggests that it is located very close to the central engine, and the very low magnetization appears to favor the formation of strong shocks, favoring shock acceleration as the dominant acceleration mechanism.

We also explored the possibility of an external-Compton (EC) emission. As there is no indication in the SED for a strong accretion-disk (or BLR) radiation field, the dominant external radiation field for which there is observational evidence, is that of the host galaxy, which, however, is likely to be distributed over (at least) kpc scales. Scaling estimates in terms of $R_{\rm ext} \equiv 1 \, R_{\rm ext, kpc}$~kpc, this implies an external radiation field of order $u_{\rm ext} \sim 10^{-11} \, R_{\rm ext, kpc}^{-2}$~erg~cm$^{-3}$. The synchrotron to $\gamma$-ray flux ratio then implies a magnetic field  of $B \sim 5 \times 10^{-6} \, R_{\rm ext, kpc}^{-1}$~G. The ratio of EC to synchrotron peak frequencies then implies the same estimate for $\gamma_e$ as for the SSC scenario above, which then, however, leads to a synchrotron peak at $\nu_{\rm sy} \sim 2 \times 10^8 \, R_{\rm ext, kpc}^{-1}$~Hz, clearly incompatible with the observed radio spectrum. We therefore conclude that an unbeamed EC model without additional (not directly constrained) external radiation fields, is implausible. 

\section{Summary}
ESO~0137$-$G007, a Fanaroff-Riley Type I \cite[][]{1974MNRAS.167P..31F} radio galaxy with a head-tail structure in a merging cluster, has added to the rich variety of misaligned AGN which have been detected in $\gamma$-rays in recent years \citep[e.g.][]{2022MNRAS.513..886B,2024ApJ...976..120P,2025ApJ...989...36P,2026arXiv260626878B}. These include Compact Symmetric Objects, other FRIs including wide-angle tailed sources, FR0s, as well as Giant Radio Galaxies. Finding a head-tail radio galaxy as a counterpart of an unidentified \gm-ray source indicates the importance of sensitive wide-field radio surveys. About a third of all known \gm-ray sources are still unidentified \citep[][]{2026arXiv260222148B}. Therefore, recent discoveries, including this work, have demonstrated that misaligned AGN might be more numerous in the \gm-ray band than previously thought \citep[][]{2022MNRAS.513..886B,2026arXiv260626878B}. Clearly, SKA will reveal many more such objects, thus improving the counterpart classification of unknown \gm-ray sources. This will also trigger high-resolution multi-wavelength follow up observations to explore the jet orientation, origin of the high-energy emission, and dynamical nature of the systems.

Although the $\gamma$-ray emission from ESO~0137$-$G007 can be understood in terms of the leptonic SSC model with electrons accelerated over a small region, the rich variety of misaligned $\gamma$-ray AGN suggests the possible co-existence of external inverse-Compton processes on different scales with photon fields from the host galaxy, extragalactic background light or the cosmic microwave background \citep[e.g.][and references therein]{2020MNRAS.491.5740P}, and possible hadronic contributions in dense fields. Our model-dependent results hints that possibly an stationary SSC model can explain the broadband SED. However, the \gm-ray emission from a few nearby misaligned AGN, such as Cen A and NGC 6251, has also been explained due to the inverse Compton scattering of the cosmic microwave background radiation. With deeper multiwavelength observations, as done for Cen A and NGC 6251, it will be possible to better constrain the \gm-ray production mechanism in ESO 0137$-$G007. High-resolution Very Long Baseline Interferometric observations will also permit us to check for superluminal motion and estimate the viewing angle accurately.

A detailed understanding of the physical processes would require deep imaging to determine the structure and extent of the radio emission, and SED modeling. The tail in ESO~0137$-$G007 has been seen to extend $>$500 kpc. While SKA-low will provide unprecedented surface-brightness sensitivity to diffuse emission even from possible earlier cycles of jet activity, SKA-mid, along with VLBI networks, will help determine the small-scale structure and help understand the origin of their $\gamma$-ray emission. The SKA will be complemented by the NewAthena X-ray observatory, the Cherenkov Telescope Array Observatory, and the neutrino telescopes IceCube-Gen2 and KM3NeT to provide a more complete understanding of $\gamma$-ray emitting misaligned AGN.

\acknowledgements

We thank the journal referee for constructive criticism. MBP gratefully acknowledges the support from the Science and Engineering Research Board (SERB), New Delhi,  under the `SERB CRG’ funding with sanction no. CRG/2023/003463. This paper includes archived data obtained through the CSIRO ASKAP Science Data Archive, CASDA (https://data.csiro.au). Based on observations obtained at the Southern Astrophysical Research (SOAR) telescope, which is a joint project of the Minist\'{e}rio da Ci\^{e}ncia, Tecnologia e Inova\c{c}\~{o}es (MCTI/LNA) do Brasil, the US National Science Foundation’s NOIRLab, the University of North Carolina at Chapel Hill (UNC), and Michigan State University (MSU). We are grateful to Drs. Denimara Dias dos Santos (LNA, Brazil) and Alberto Rodr\'iguez Ardila (ON, Brazil) for granting access to their observing time on SOAR to observe this source. SP is supported by the international Gemini Observatory, a program of NSF NOIRLab, which is managed by the Association of Universities for Research in Astronomy (AURA) under a cooperative agreement with the U.S. National Science Foundation, on behalf of the Gemini partnership of Argentina, Brazil, Canada, Chile, the Republic of Korea, and the United States of America.

\bibliographystyle{aasjournal}
\bibliography{Master}

\end{document}